\documentclass[showpacs,amsmath,amssymb,superscriptaddress,floatfix,reprint,prl]{revtex4-2}
\usepackage{blindtext}
\usepackage{amsfonts}
\usepackage{mathrsfs}
\usepackage{amsmath}
\usepackage{graphicx}
\usepackage{cases}
\usepackage{amsthm}
\usepackage{amssymb}
\usepackage{booktabs}
\usepackage{multirow}
\usepackage{makecell}
\usepackage{color}
\usepackage{siunitx}
\usepackage{setspace}
\usepackage[colorlinks=true, citecolor=blue, linkcolor=blue, urlcolor=blue]{hyperref}
\definecolor{coral}{RGB}{255,127,80}
\newcommand{\ket}[1]{\mbox{$|#1\rangle$}}

\newcommand{\blue}[1]{\textcolor{blue}{#1}}

\begin{document}

\title{Quantum photonic frequency processor on thin-film lithium niobate}

\author{Ran Yang}
\thanks{These authors contribute equally to this work.}
\affiliation{National Laboratory of Solid State Microstructure, School of Physics, School of Electronic Science and Engineering, Jiangsu Key Laboratory of Quantum Information Science and Technology, and Collaborative Innovation Center of Advanced Microstructures, Nanjing University, Nanjing 210093, China}

\author{Wei Zhou}
\thanks{These authors contribute equally to this work.}
\affiliation{National Laboratory of Solid State Microstructure, School of Physics, School of Electronic Science and Engineering, Jiangsu Key Laboratory of Quantum Information Science and Technology, and Collaborative Innovation Center of Advanced Microstructures, Nanjing University, Nanjing 210093, China}

\author{Dong-Jie Guo}
\thanks{These authors contribute equally to this work.}
\affiliation{National Laboratory of Solid State Microstructure, School of Physics, School of Electronic Science and Engineering, Jiangsu Key Laboratory of Quantum Information Science and Technology, and Collaborative Innovation Center of Advanced Microstructures, Nanjing University, Nanjing 210093, China}

\author{Hong-Ming Ke}
\affiliation{National Laboratory of Solid State Microstructure, School of Physics, School of Electronic Science and Engineering, Jiangsu Key Laboratory of Quantum Information Science and Technology, and Collaborative Innovation Center of Advanced Microstructures, Nanjing University, Nanjing 210093, China}

\author{Linrunde Tao}
\affiliation{National Laboratory of Solid State Microstructure, School of Physics, School of Electronic Science and Engineering, Jiangsu Key Laboratory of Quantum Information Science and Technology, and Collaborative Innovation Center of Advanced Microstructures, Nanjing University, Nanjing 210093, China}

\author{Ying Wei}
\affiliation{National Laboratory of Solid State Microstructure, School of Physics, School of Electronic Science and Engineering, Jiangsu Key Laboratory of Quantum Information Science and Technology, and Collaborative Innovation Center of Advanced Microstructures, Nanjing University, Nanjing 210093, China}

\author{Jia-Chen Duan}
\affiliation{National Laboratory of Solid State Microstructure, School of Physics, School of Electronic Science and Engineering, Jiangsu Key Laboratory of Quantum Information Science and Technology, and Collaborative Innovation Center of Advanced Microstructures, Nanjing University, Nanjing 210093, China}

\author{Yu Cui}
\affiliation{National Laboratory of Solid State Microstructure, School of Physics, School of Electronic Science and Engineering, Jiangsu Key Laboratory of Quantum Information Science and Technology, and Collaborative Innovation Center of Advanced Microstructures, Nanjing University, Nanjing 210093, China}

\author{Kunpeng Jia}
\affiliation{National Laboratory of Solid State Microstructure, School of Physics, School of Electronic Science and Engineering, Jiangsu Key Laboratory of Quantum Information Science and Technology, and Collaborative Innovation Center of Advanced Microstructures, Nanjing University, Nanjing 210093, China}

\author{Zhenda Xie}
\affiliation{National Laboratory of Solid State Microstructure, School of Physics, School of Electronic Science and Engineering, Jiangsu Key Laboratory of Quantum Information Science and Technology, and Collaborative Innovation Center of Advanced Microstructures, Nanjing University, Nanjing 210093, China}
\author{Zhongjin Lin}
\email{linzhj56@mail.sysu.edu.cn}
\affiliation{State Key Laboratory of Optoelectronic Materials and Technologies, School of Electronics and Information Technology, Sun Yat-sen University, Guangzhou 510275, China}

\author{Xinlun Cai}
\affiliation{State Key Laboratory of Optoelectronic Materials and Technologies, School of Electronics and Information Technology, Sun Yat-sen University, Guangzhou 510275, China}
\affiliation{Hefei National Laboratory, Hefei 230088, China}

\author{Yan-Xiao Gong}
\email{gongyanxiao@nju.edu.cn}
\affiliation{National Laboratory of Solid State Microstructure, School of Physics, School of Electronic Science and Engineering, Jiangsu Key Laboratory of Quantum Information Science and Technology, and Collaborative Innovation Center of Advanced Microstructures, Nanjing University, Nanjing 210093, China}
\affiliation{Hefei National Laboratory, Hefei 230088, China}

\author{Shi-Ning Zhu}
\affiliation{National Laboratory of Solid State Microstructure, School of Physics, School of Electronic Science and Engineering, Jiangsu Key Laboratory of Quantum Information Science and Technology, and Collaborative Innovation Center of Advanced Microstructures, Nanjing University, Nanjing 210093, China}

\begin{abstract}
\textbf{The rapid development of photonic quantum information processing necessitates precise and programmable control over optical frequency, a capability critical not only for achieving photon indistinguishability but also for exploiting a virtually unbounded frequency dimension.
%exploiting virtually unbounded frequency dimension to encode quantum states. %While frequency manipulation of photons has been achieved with commercial travelling wave electro-optic modulators and third-order nonlinear optical processes, 
However, efficient and scalable processing of frequency-encoded photon states remains challenging, primarily due to the limited nonlinear optical interaction in most photonic materials. %electro-optic interaction in the majority of integrated photonic materials. %, which hinders their quantum applications requiring active manipulation of photons. 
Here, by harnessing the high-performance thin-film lithium niobate electro-optic (EO) platform, we demonstrate an integrated quantum photonic frequency processor that enables coherent and programmable control of photon frequency with high precision. 
We establish a scalable architecture for frequency-encoded quantum information processing. % towards  circuits. 
Using a fully integrated photonic chip, we realize a universal set of frequency-encoded quantum logic gates, including arbitrary single-qubit rotation gates and the two-qubit controlled-phase gate. Furthermore, we demonstrate its application in high fidelity characterization of frequency-bin entangled states. Our work reveals the unprecedented potential of utilizing the frequency degree of freedom in integrated quantum photonic systems.
%for integrated utilization of the frequency degree of freedom in quantum photonic systems.
%The results advance integrable frequency-based quantum processor into 
}
\end{abstract}

\maketitle
\noindent\textbf{Introduction}

\noindent The frequency degree of freedom (DOF) plays a crucial role in optical information processing across both classical \cite{fortier201920} and quantum \cite{kues2019quantum} regimes. As an inherently continuous DOF, optical frequency can be discretized to create ultrahigh-dimensional quantum state spaces in a single photon or optical mode, offering a powerful route towards scalable quantum systems. Significant advances have been made in the generation of both large-scale multipartite continuous-variable \cite{PhysRevLett.112.120505,roslund2014wavelength,cai2017multimode,jia2025continuous,roh2025generation}, and high-dimensional photonic frequency-bin entangled states \cite{doi:10.1126/science.aad8532,kues2017chip,mahmudlu2023fully,xie2015harnessing,PhysRevA.82.013804,reimer2019high}. However, linear optics confines operations on the frequency DOF to passive separating, mixing, and phase adjustment of frequency modes, preventing genuine cross-mode operations that typically require nonlinear optics. Consequently, the scope of frequency-encoded quantum information processing is more constrained in comparison to more versatile DOFs such as polarization, spatial or path. Hence, to unlock the potential of frequency-encoded quantum technologies, it is crucial to develop fully scalable, robust, and reconfigurable photonic frequency control devices \cite{lu2023frequency,dutt2024nonlinear}.
%However, within the framework of linear optics, operations on the frequency degree of freedom (DOF) are fundamentally confined to passive separating, mixing, and phase adjustment of frequency modes, without enabling interactions between distinct frequency modes. 
%In continuous variable~(CV) quantum information, the multitude of modes in an optical frequency comb facilitates the generation of large-scale multi-partite entangled states \cite{PhysRevLett.112.120505,roslund2014wavelength,cai2017multimode,jia2025continuous,roh2025generation}, quantum enhanced  spectroscopy \cite{doi:10.1126/science.ads6292,plh2-cr8s,wan2025quantum,shi2023entanglement}, and boson sampling~\cite{borghi2025bipartite}. While in discrete variable~(DV) regime, a frequency-bin qudit can potentially enable the encoding of a large amount of noise-resilient quantum information on a single photon. Extensive efforts have been made in the manipulation and applications of photonic photonic frequency-bin quantum states, %leading to many important achievements, for example, in 
%such as high-dimensional entanglement generation~\cite{doi:10.1126/science.aad8532,kues2017chip,mahmudlu2023fully,xie2015harnessing,PhysRevA.82.013804,reimer2019high}, quantum logic gates~\cite{lukens2016frequency,PhysRevLett.120.030502,lu2019controlled}, multi-mode quantum storage~\cite{PhysRevLett.123.080502,PhysRevLett.113.053603,yang2018multiplexed,doi:10.1126/sciadv.abo4538}, and multiplexed quantum networks~\cite{khodadad2025frequency,tagliavacche2025frequency}.

The fast-driven electro-optic modulation (EOM), typically based on lithium niobate (LN)~\cite{parriaux2020electro,hu2025integrated}, provides a convenient and programmable means to perform fast, clean, and low-loss spectral operations by utilizing standard radio-frequency and microwave electronics. Combined with low-loss waveguides, and high-$Q$ microresonators, high-performance photonic frequency manipulation devices have been developed, such as frequency shifters and beam splitters~\cite{zhang2019electronically,buddhiraju2021arbitrary,hu2021chip,kapoor2025electro}, comb generators~\cite{hu2022high}, frequency circulation~\cite{herrmann2022mirror}, and spectral control~\cite{karpinski2017bandwidth,zhu2022spectral,sosnicki2023interface}. However, when applied to quantum logic gates and computation, the inherent infinite frequency sidebands generated by EOM modulations would extend beyond the designated computation Hilbert spaces, thereby limiting the success probability and fidelity of the logic operations. To address this problem, a framework for scalable frequency-encoded quantum information processing was proposed by cascading pulse-shaper/EOM modules~\cite{lukens2016frequency}. Despite successful demonstrations of high-fidelity beam splitters and tritters with a single module~\cite{PhysRevLett.120.030502}, realizing a complex logic gate like two-qubit controlled-phase gate in this framework requires several cascaded modules, which increase circuit depth, accumulate optical loss, and ultimately limit the fidelity and scalability~\cite{lu2019controlled}. Consequently, the realization of more efficient, scalable, and high-fidelity frequency-encoded quantum information-processing architectures remains an outstanding challenge.

\begin{figure*}[!t]
  \centering
  \includegraphics[scale = 1]{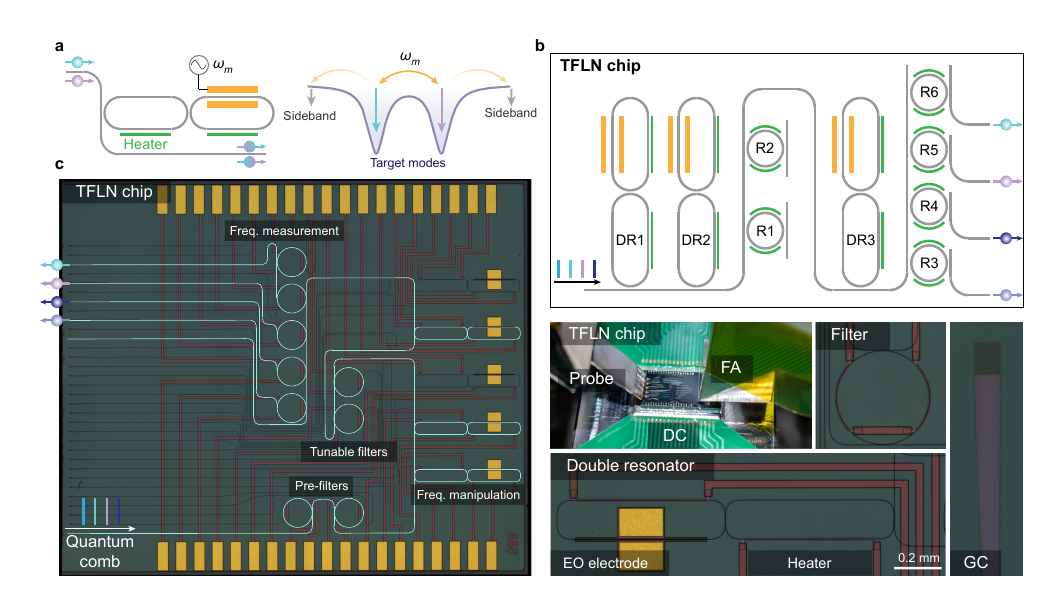}
  \caption{\textbf{Description of the Quantum Photonic frequency processor chip.} \textbf{a}, Schematic and operating principle of microwave-driven coupled double resonators (DR). %The higher-order sidebands are suppressed due to the near zero intra-cavity state density. %versatile quantum operations in frequency domain. 
  \textbf{b}, Schematic diagram of the chip design. %The photonic integrated circuit incorporating double resonators, microring filters, and electrodes for reconfigurable quantum operations. %to implement multifunctional tasks with photonic frequency qubits. 
  \textbf{c}, Optical microscope images of the chip and the test setup. FA, fiber array; \blue{DC, direct-current source; R, microring resonator;} GC, grating coupler.
  }
  \label{fig:schematic}
\end{figure*}

Here we present a monolithic and reconfigurable quantum photonic frequency processor chip fabricated on thin-film lithium niobate (TFLN), a promising platform for integrated photonics~\cite{zhu2021integrated}. A key module is the microwave-driven coupled double resonators (DR), which can realize a tunable beam splitter between two frequency bins with minimal inter-bin crosstalk. The Hong-Ou-Mandel interference between two photons in distinct frequency bins is observed with a visibility of $(94.9\pm1.8)\%$. By defining frequency-bin encoded Hilbert spaces, we demonstrate high-quality arbitrary single-qubit rotation logic gates with an average fidelity of $(97.1\pm0.6)\%$. Then for the first time, we experimentally realize the canonical ancilla-free linear optical controlled-phase~(or equivalently, controlled-Not) gate~\cite{CP2002,CNOT2002} utilizing frequency-bin encoding, in contrast to previous implementations on other DOFs, such as path~\cite{o2003demonstration,politi2008silica}, polarization~\cite{CNOTPL1,CNOTPL2,CNOTPL3}, waveguide mode~\cite{CNOTmode2021}. The fidelity of this gate is measured to be $(91.4\pm1.4)\%$ that could be improved to {$98.9 \%$}, provided a higher quality photon source. Our approach paves the way for quantum information processing in the continuous frequency DOF, as well as shows a promising way for the seamless integration of frequency-nonidentical photons.  

\vspace{0.5em}

\noindent\textbf{Results}

\noindent\textbf{Scheme and device}

\noindent %The basic module of our integrated optical frequency processor is a coupled DR, as illustrated in Fig.~\ref{fig:schematic}a, with its working principle shown beside. . %illustrates the working principle of the basic device to manipulate optical frequency. 
% We utilize the avoided mode crossing~(AMX) in coupled double resonators to break the uniform resonance spacing. 
A schematic of the basic module, a coupled DR, along with its operating principle, is presented in Fig.~\ref{fig:schematic}a. When one resonator is driven by a microwave signal with a frequency close to the optical mode splitting of the DR, coherent interconversion between the two frequency modes is induced by the EO effect. Higher-order sidebands are intrinsically suppressed due to the near-zero density of intra-cavity states at those frequencies. This device functions as a frequency-bin beam splitter~(f-BS) acting on two input and two output frequency bins, analogous to a spatial beam splitter \blue{(see Supplementary Information Sec. I).} Its operation can be described by the unitary matrix $U$ of a standard beam splitter with transmissivity $T$ and reflectivity $R$, scaled by a total efficiency $\eta$ (for details, see Supplementary Information Sec. II). Consequently, by controlling the amplitude and phase of microwave driving signals, versatile quantum operations on frequency-bin encoded qubits can be realized.

\begin{figure*}[t]
  \centering
  \includegraphics[scale = 1]{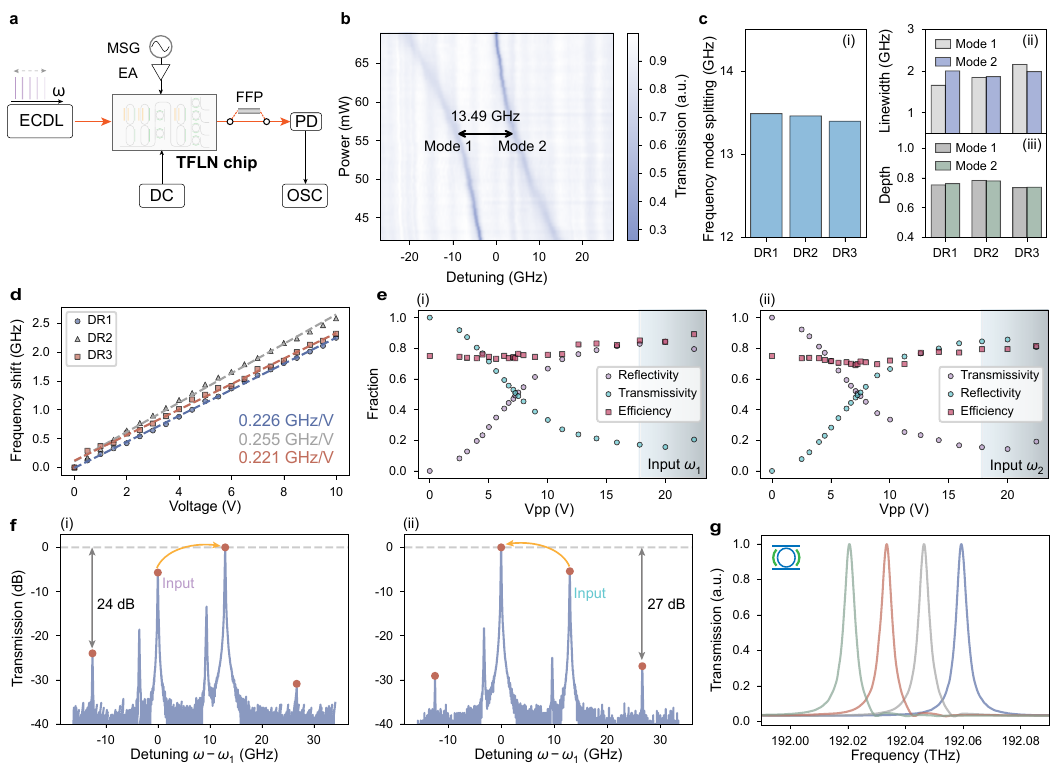}
  \caption{\textbf{Chip characterization.} \textbf{a}, Experimental setup for characterizing the chip. ECDL, external cavity diode laser; MSG, microwave signal generator; EA, electrical amplifier; FFP, fiber Fabry-P\'erot cavity; PD, photodetector; %DC, direct current source; 
  OSC, oscilloscope. 
  \textbf{b},~
  Transmission spectrum of DR1 against the thermal tuning power. The minimum frequency mode splitting is 13.49 GHz, with balanced transmission of the two modes.
  %Thermal response of DR1, showing a frequency mode splitting of 13.49 GHz. 
  \textbf{c}, Performance of three DRs, including (i) frequency mode splitting, (ii) linewidth, and (iii) transmission-dip depth. 
  \textbf{d}, The resonance shift under DC voltages, with the corresponding EO-response values, obtained from linear fits, listed at lower right corner.  
  \textbf{e}, Splitting ratio and total efficiency $\eta$ of the f-BS for DR1 with input $\omega_1$ (i) and $\omega_2$ (ii). 
  %\textbf{f}, Frequency diagram to measure the response of f-BSs. The input frequencies $\omega_1$ and $\omega_2$ are symmetrically detuned with respect to the DR mode doublet. 
  \textbf{f}, Frequency distribution at maximal interconversion efficiency, measured by scanning the resonant frequency of a FFP cavity (FSR, 10 GHz). %The suppression of high order side-band beyond $\omega_1$ and $\omega_2$ is estimated to exceed 24 dB. The two unlabeled peaks come from another polarization mode of FFP.
  \textbf{g}, Transmission spectra of the add-drop microring filters.
  }
  \label{fig:Fig_2}
\end{figure*}

%We demonstrate an integrated photonic frequency encoded quantum state processor (Fig.~\ref{fig:schematic}b). It consists of several basic components, including DRs, microring resonators~(MRR), and grating couplers. 
The schematic diagram of our integrated quantum photonic frequency processor is shown in Fig.~\ref{fig:schematic}b. %In addition to DRs, the chip consists of several other basic components, including microring resonators~(MRRs), and grating couplers \blue{(GCs)}. %The DRs are employed for versatile quantum operations of frequency qubit, by controlling the amplitude and phase of microwave driving signals. 
In addition to three DRs, two MRRs, R1 and R2, are used to implement frequency-dependent attenuation in the canonical ancilla-free linear optical controlled-phase gate. Four MRRs, R3, R4, R5, and R6, are employed to separate frequency bins and perform frequency-bin projective measurements. 
%\red{[efficiency]}ensuring stable and low-loss optical interfacing. 
Optical microscope images of the integrated photonic chip are shown in Fig.~\ref{fig:schematic}c, which is fabricated based on the {6-inch} wafer-scale TFLN technology~(see Methods).  Grating couplers are used to  couple light between on-chip waveguides and fibers with an efficiency of $25\%$ and a bandwidth of 2.5 THz. Two pre-filters are employed to eliminate photons at unwanted frequencies. Note that, there are two other DRs are on the chip, which are detuned far from the frequencies used in our experiments. % In our experiments, we choose three out of the five DRs to be activated, which are named DR1, DR2, and DR3, respectively. The other DRs are detuned far from the frequencies used in our experiments to minimize unwanted couplings and losses.
%Note that only DR1, DR2, and DR3 are activated in our experiments, whereas DR3 and DR5 are detuned far from the frequencies used in our experiments to minimize unwanted couplings and losses.

% \begin{table*}[t]
% 	\centering
% 	\begin{flushleft}
%     TABLE. \uppercase\expandafter{\romannumeral1}. Comparison of on-chip manipulation of photonic states in various degree of freedom.
%     \end{flushleft}
% 	\begin{tabular}{ccccccccc}
%     111111
%     \end{tabular}	
% 	\label{tab:comparison}
% \end{table*}

\vspace{0.5em}
\noindent\textbf{Chip characterization}

\noindent %Before investigating the system-level functionalities, 
We first characterize the performance of each on-chip building block. 
The experimental setup is illustrated in Fig.~\ref{fig:Fig_2}a. The optical spectroscopy of the resonators is performed by sweeping an external-cavity diode laser (ECDL) across the resonance frequencies. A microwave signal from a microwave signal generator (MSG) is amplified and then applied to the electrodes of DRs to drive the tunable f-BS. The EO responses of the f-BSs are characterized via on-chip MRR filters that selectively separate the frequency components of the output light. A narrow-linewidth fiber Fabry-P\'erot (FFP) cavity \blue{with a full width at half maximum (FWHM) of 18 MHz} is used to resolve and characterize the higher-order sidebands in high spectral resolution. The output optical signal is detected by a \blue{photodetector} (PD) and recorded by an \blue{oscilloscope} (OSC) for further analysis.

%\begin{figure*}[!t]
%  \centering
%  \includegraphics[scale = 1]{figs/Fig_3.pdf}
%  \caption{\textbf{Frequency domain photonic interferences.} \textbf{a}, Experiment configuration of frequency domain Mach-Zehnder interference. EPS, \blue{eletrical} phase shifter. \textbf{b}, Characterization of f-MZI using CW light. \textbf{c}, Demonstration of f-MZI using heralded single photon.  \textbf{d}, Schematic of frequency domain Hong-Ou-Mandel interference. TCSPC, time-correlated single photon counting. \textbf{e}, Histogram of time-correlated coincidence counts. The curves are fitted using a convolution of a two-sided exponential decay function, representing the two-photon temporal correlation, with a 512 ps time-resolution window. \textbf{d}, Hong-Ou-Mandel interference at different beam splitting ratio. 
%  }
%  \label{fig:Fig_3}
%\end{figure*}

% The integrated photonic chip is fabricated based on a \red{4-inch} wafer-scale TFLN technology (see Methods), as displayed in Fig.~\ref{fig:Fig_2}b. The TFLN photonic chip consists of several basic devices, including three optical molecules, six microrings, and grating couplers.
By thermally tuning the resonance of the resonator directly coupled to bus waveguide of DR1, we measure the behavior of frequency transmission against thermal power, as shown in Fig.~\ref{fig:Fig_2}b. At the transmission-balanced point, the frequency mode splitting $2g$ is estimated to be $13.49$~GHz, which is close to the designed 12.95 GHz that matches the frequency spacing of our quantum comb photon-pair source (see Supplementary Information Sec. III. B). The three DRs, DR1, DR2, and DR3, exhibit consistency in frequency mode splitting, resonance linewidth, and dip transmission, as shown in Fig.~\ref{fig:Fig_2}c~(i), (ii), and (iii), respectively. Their resonance frequency shifts as a function of a direct-current voltage applied to the inner resonator are presented in Fig.~\ref{fig:Fig_2}d, with the corresponding EO-response values estimated to be 0.226, 0.255, and 0.222~$\text{GHz/V}$, respectively. These modulation efficiencies can be further improved with optimized electrode designs. 

To characterize the splitting ratio of the f-BS driven by a microwave signal set at frequency $\omega_m=12.95$~GHz, we input a continuous wave~(CW) laser at frequencies $\omega_1=192.02052$~THz and $\omega_2=192.03347$~THz, which are symmetrically detuned with respect to the DR doublet shown in Fig.~\ref{fig:schematic}a. The measured results of reflectivity and transmissivity for input frequencies $\omega_1$ and $\omega_2$ under different voltages are presented in Fig.~\ref{fig:Fig_2}e~(i) and~(ii), respectively. 
%The results of frequency interconversions are shown in Fig.~\ref{fig:Fig_2}g~(i) and~(ii), where the shaded areas indicate the gradual saturation of the electrical amplifiers (EAs). 
Note that the shaded areas indicate the gradual saturation of the electrical amplifiers (EAs). {The total efficiency is estimated to exceed 69$\%$, and could be further improved to 97$\%$ through optimization of the intrinsic waveguide loss~\cite{gao2022lithium,zhu2024twenty}.} The higher-order sideband suppression ratios are observed to exceed 24~dB at maximal conversion efficiencies, as shown in Fig.~\ref{fig:Fig_2}f~(i) and~(ii). {This intrinsic suppression ensures closed, low-leakage operations within a well-defined two-frequency-bin subspace, a key requirement for scalable quantum circuits.}
% , confirming the closed operations inside the subspace composed of two frequency-bins. 
{The two unlabeled peaks come from resonant modes associated with another orthogonal polarization in the FFP cavity.} The microring filters are designed with a free spectral range~(FSR) of 100 GHz, which in principle support an unambiguous discrimination of four frequency-bins spanning $3\times12.95$~GHz involved in our quantum experiments. The measured linewidths of the filters are approximately 4 GHz~(Fig.~\ref{fig:Fig_2}g), ensuring measurement crosstalk between nearest bins less than $3$\%. The on-chip efficiency of the filters is estimated at over $94.6$\%.

%We show that the individual elements of the proposed photonic chip provide high and consistent performance.

%\begin{figure*}[!t]
%  \centering
%  \includegraphics[scale = 1]{figs/Fig_4.pdf}
%  \caption{\textbf{Frequency qubits manipulation.} \textbf{a}, Encoding of two frequency qubits. \textbf{b}, Chip configuration for characterization of frequency encoded two photon CZ gate. \textbf{c}, Measured probability distributions of CZ gate with different input states. \textbf{d}, Generation of frequency-bin encoded entangled state. \textbf{e}, Chip configuration for characterization of frequency-bin entangled state. \textbf{f}, Measured entanglement curves.
%  }
%  \label{fig:Fig_4}
%\end{figure*}

\begin{figure}[t]
  \centering
  \includegraphics[scale = 1]{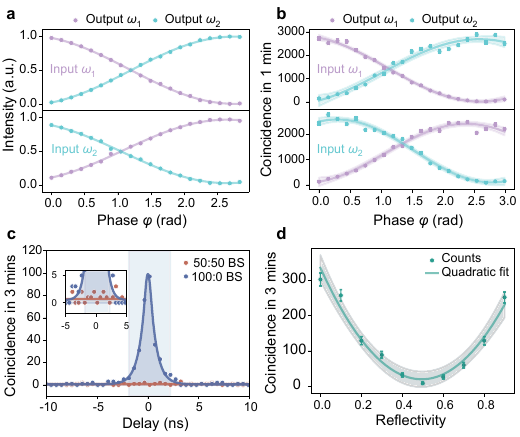}
  \caption{\textbf{Frequency domain photonic interferences.} \textbf{a}, Characterization of f-MZI using CW light. \textbf{b}, Demonstration of f-MZI using heralded single photon. \textbf{c}, Histogram of time-correlated coincidence counts. \textbf{d}, Hong-Ou-Mandel interference at different reflectivities. 
  }
  \label{fig:Fig_3}
\end{figure}

\vspace{0.5em}
\noindent\textbf{Frequency-domain photonic interferences}

\noindent{The operation of the f-BS on single photons can be modeled as a unitary matrix $U$ within a two-dimensional Hilbert space spanned by two frequency bins $\{\ket{\omega}, \ket{\omega + \Delta\omega}\}$ with a spacing of $\Delta\omega = 12.95$ GHz. Due to the bandwidth constraints of the grating couplers, these frequency bins can be selected across a wide operational range from $190.1734$ THz to $192.6459$ THz. The transformation is expressed as:}
\begin{equation}
U = \begin{bmatrix} \sqrt{T} & e^{i\theta}\sqrt{R} \\ -e^{-i\theta}\sqrt{R} & \sqrt{T} 
\end{bmatrix},
\end{equation}
where $R$ and $T$ represent the reflectivity and transmissivity, and $\theta$ is the relative phase controlled by the microwave drive. By tuning these parameters, the f-BS essentially acts as a reconfigurable single-qubit gate for frequency-encoded states.

To evaluate the functional performance of the TFLN photonic frequency manipulation system, we first validate the coherence of the f-BS by utilizing a frequency-domain Mach-Zehnder interferometer~(f-MZI) configured for two specific frequency bins, $\omega_1$ and $\omega_2$, with both classical laser light and single-photon state inputs (for details, see Supplementary Information Sec. I). We select DR1 and DR3 shown in Fig.~\ref{fig:schematic}b to serve as the two f-BSs in the f-MZI, with the resonances of all other DRs thermally tuned away from the operational frequencies. The resonances of DR1 and DR3 are initialized to be identical, both operating at a $50:50$ splitting ratio. The phase difference between the two arms in the f-MZI is determined by the relative phase $\varphi$ between two microwave signals, which is tuned by an electrical phase shifter. The two-frequency-bin light is spectrally separated by two MRRs and detected  using off-chip power meters. 

%Classically, b
By injecting a CW laser at frequencies $\omega_1$ and $\omega_2$, the normalized output intensities at two frequency ports are shown in Fig.~\ref{fig:Fig_3}a, which presents high average visibility $\mathcal{V}_\text{ave} = \langle({I}_\text{max} - {I}_\text{min})/({I}_\text{max} + {I}_\text{min})\rangle = 97.2\%$ for four curves. 
% In quantum experiments

\begin{figure*}[!t]
  \centering
  \includegraphics[scale = 1]{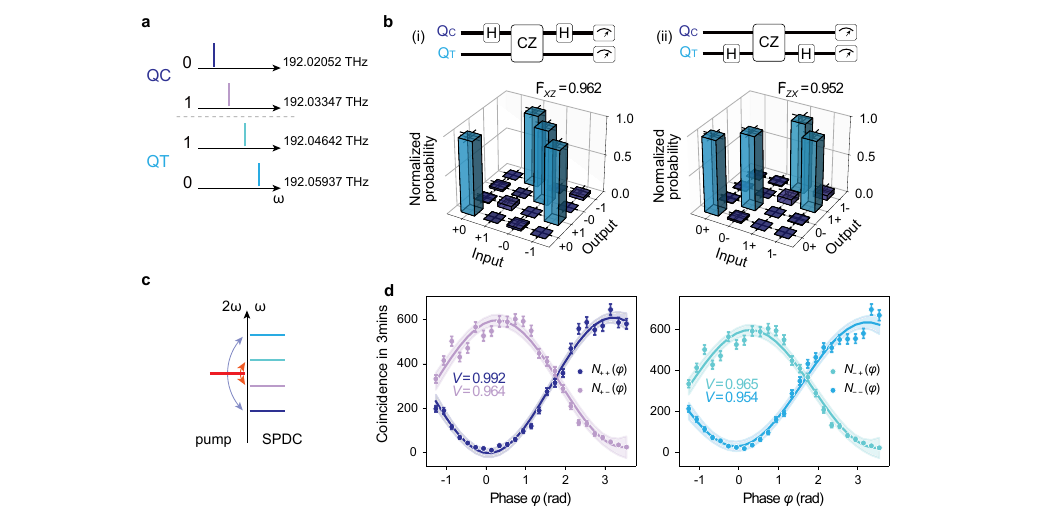}
  \caption{\textbf{Frequency qubits manipulation.} \textbf{a}, Encoding of qubits in photonic frequency-bins. \textbf{b}, Measured probability distributions of CZ gate with different input states. \textbf{c}, Generation of frequency entangled state through SPDC. \textbf{d}, Measured entanglement curves.
  }
  \label{fig:Fig_4}
\end{figure*}

The heralded single photons centered at frequencies $\omega_1$ and $\omega_2$ with a Lorentzian bandwidth of 202 MHz are generated from the spontaneous parametric down-conversion (SPDC) inside a monolithic cavity with a FSR of 12.95~GHz (see Supplementary Information Sec. III. B). After passing through the f-MZI, the output photons are routed to off-chip superconducting nanowire single-photon detectors (SNSPDs). The coincidence counts are recorded by a time-correlated single photon counting (TCSPC) module. %, ensuring high detection efficiency \red{($\approx 95\%$ [verification])} and low dark-count rates \red{($\approx$ 100 cps [verification])}.}
The measured results are presented in Fig.~\ref{fig:Fig_3}b, showing a high average visibility of $\mathcal{V}_\text{Q}=97.1\pm0.6\%$. This result confirms the ability to achieve high-fidelity and coherent control of frequency-bin photonic states.% Such operations constitute a universal set for single-qubit gates \blue{in Hilbert space spanned by states $\ket{\omega_1}$ and $\ket{\omega_2}$}.

Then, we perform the Hong-Ou-Mandel (HOM) interference, a fundamental building block in photonic quantum information processing, between two photons in distinct frequency bins. 
% The chip configuration is illustrated in Fig.~\ref{fig:Fig_3}d. 
A pair of photons at frequencies $\omega_1$ and $\omega_2$ generated from SPDC are coupled into the TFLN chip. We employ DR3 as the f-BS in the HOM interferometer for its widest splitting-ratio range among the three DRs (see Fig. S9 in Supplementary Information Sec. III). After frequency separation, the output photons at $\omega_1$ and $\omega_2$ are directed to two separate SNSPDs for detection. Then the coincidence counts are recorded with a time-resolution window of 512 ps.
% \zw{Coincidence counts between the two frequency channels are measured at different splitting ratio of the frequency BS, which is tuned by varying the power of the microwave driving signal.}
%The coincidence counts are recorded with a 512 ps time-resolution window
% recorded by a time-correlated single photon counting (TCSPC) module and 
%measured at different reflectivities of the f-BS, which are tuned by varying the microwave driving power. 
Figure~\ref{fig:Fig_3}c shows the time-correlated coincidence counts with the splitting ratio set at 100:0 and 50:50, respectively. The curves are fitted using a convolution of a two-sided exponential decay function. %representing the two-photon temporal correlation. 
The result clearly reveals the photon bunching phenomenon in the HOM interference at balanced splitting condition. The visibility of the HOM interference is calculated as $\mathcal{V}_\text{HOM} = ({N}_\text{max}-{N}_\text{min})/{N}_\text{max} = 94.9\pm1.8\%$ without background count subtraction, where ${N}_\text{max}$ and ${N}_\text{min}$ denote maximal and minimal counts, respectively. We further measure the HOM interference at different reflectivities of the f-BS, with the experiment results presented in Fig.~\ref{fig:Fig_3}d. 
%The relation of coincidence counts versus the f-BS reflectivity is presented in Fig.~\ref{fig:Fig_3}d. 
These results confirm the capability of the chip for high-visibility nonclassical interference between two photons in distinct frequency bins.

\vspace{0.5em}
\noindent\textbf{Frequency-bin encoded qubits control}

\noindent %Building upon the demonstrated high-performance frequency manipulation capability, we further employ our TFLN devices to realize quantum control of photonic frequency-encoded qubits. 
As illustrated in Fig.~\ref{fig:Fig_4}a, four frequency-bins are encoded as the control ($\text{Q}_\text{C}$) and target ($\text{Q}_\text{T}$) qubits. The chip is configured for implementation and characterization of the ancilla-free linear optical controlled-phase gate~\cite{CP2002,CNOT2002}, or called  the CZ gate (see Supplementary Information Sec. II. C). In the circuit, DR2 is tuned to achieve a splitting ratio of $T:R=1:2$ for frequency bins encoded as $\ket{1}_\text{C}$ and $\ket{0}_\text{T}$. The MRRs R1 and R2 are tuned to introduce an  attenuation of $2/3$ for photons in state $\ket{0}_\text{C}$ and $\ket{1}_\text{T}$, respectively.  %By controlling the detuning of the MRRs with respect to the frequency-bins, R1 and R2 are tuned to introduce a $2/3$ attenuation for photons in state $\ket{0}_\text{C}$ and $\ket{1}_\text{T}$, respectively. 
Under this configuration, the frequency-bin encoded CZ gate is constructed. The state preparation is implemented by DR1. The frequency-bin state projection measurements are realized by DR3 and the subsequent MRRs, R3, R4, R5, and R6. 
We measure the input-output statistics in Pauli basis $XZ(\{\ket{\pm}_c,{\ket{0/1}_t}\}$ (Fig.~\ref{fig:Fig_4}b(i)) and $ZX(\{\ket{0/1}_c,\ \ket{\pm}_t\}$ (Fig.~\ref{fig:Fig_4}b(ii)), where DR1 and DR3 are configured to perform the Hadamard operation on either control or target qubit. The lower bound of the  quantum process fidelity of the gate can be estimated as $\mathcal{F} \ge (\mathcal{F}_{XZ} + \mathcal{F}_{ZX}-1) = (91.4\pm1.4)\%$ \cite{PhysRevLett.94.160504}, where $\mathcal{F}_{XZ}$ and $\mathcal{F}_{ZX}$ are the fidelity of measured input-output probabilities. The deviation mainly stems from the low coincidence-to-accidental ratio. With a higher quality photon source, the fidelity could be improved to $98.9 \%$ (for details, see Supplementary Information Sec. II. C). %can be attributed to various factors, including inaccuracies in state preparation, projective measurements, and low coincidence accidental ratios. 

We then utilize the device to characterize two photon frequency-bin entanglement. The entangled  state {$\left(\ket{00}+\ket{11}\right)/\sqrt{2}$} is generated from a single-tone CW pumped SPDC process (Fig.~\ref{fig:Fig_4}c), where only photons in four selected bins are injected into TFLN chip using off-chip filters. By tuning the phase $\varphi$ of the microwave signal on DR2, the chip is configured to perform projective measurements on superposition states $\left(\ket{0}\pm\ket{1}\right)/\sqrt{2}$ and $\left(\ket{0}\pm e^{-i\varphi}\ket{1}\right)/\sqrt{2}$ at DR1 and DR2 for two frequency qubits, respectively. The measured coincidence counts for $N_{++}(\varphi)$, $N_{+-}(\varphi)$, $N_{-+}(\varphi)$, and $N_{--}(\varphi)$ are shown in Fig.~\ref{fig:Fig_4}d. The observed average interference visibility of $96.9\pm0.6\%$ confirms that frequency-encoded entanglement is preserved with minimal decoherence, highlighting the multifunctional capabilities of our TFLN quantum photonic platform for coherent frequency manipulation.

\vspace{0.5em}
\noindent\textbf{Conclusions}

%In this work, we have successfully demonstrated a reconfigurable photonic chip towards integrated quantum information processing of frequency-encoded qudits. 

\noindent 
The ability to coherently manipulate photons in the frequency domain provides a new dimension for scalable quantum information processing. In this work, we have developed an integrated quantum photonic frequency processor on  TFLN platform, uniting high-speed EOM with the scalability of integrated photonics. This platform enables programmable frequency-domain operations that serve as universal building blocks for quantum logic, establishing a practical route toward reconfigurable and high-dimensional quantum circuits. Although the CZ gate realized here is non-deterministic, its demonstrated functionalities can be further exploited in more advanced protocols for both continuous-variable and discrete-variable regimes. The frequency processor lays the groundwork for quantum state engineering, spectral multiplexing, and hybrid encoding schemes that coherently integrate frequency, spatial, and polarization modes within a single chip-scale architecture.

%The integration of strong electro-optic interactions, low optical loss, and CMOS compatibility marks an important step toward large-scale quantum photonic processors. Beyond discrete quantum logical operations, the frequency processor demonstrated here lays the groundwork for broadband quantum state engineering, spectral multiplexing, and hybrid encoding schemes that coherently link frequency, spatial, and polarization modes within a single chip-scale architecture.

In the future, the heterogeneous integration of quantum light sources \cite{xue_ultrabrightmultiplexed_2021,ma_ultrabrightquantum_2020,javid_ultrabroadbandentangled_2021}, frequency processors, and single-photon detectors \cite{lomonte2021single,prencipe2023wavelength} will enable fully self-contained quantum photonic systems \cite{psiquantum2025manufacturable} capable of on-chip quantum state generation, manipulation, and measurement. Such advances will not only eliminate coupling losses and enhance stability, but also realize dense frequency-multiplexed quantum circuits for communication and computation. The continued convergence of integrated electro-optics, nonlinear optics, and quantum state engineering is expected to transform the frequency degree of freedom from a passive carrier into an active, programmable resource for universal quantum information technologies. Moreover, as LN waveguides are well-suited for developing quantum integrated circuits \cite{jin2014chip,n2y3-2bmz}, our approach hold great promise integrated in quantum photonic chips.
% (to be fixed).

% \red{[source integration]}

\vspace{1.5em}
\noindent\textbf{Acknowledgments}
\begin{acknowledgments}
\noindent This research is supported by Quantum Science and Technology-National Science and Technology Major Project (Grant Nos. 2021ZD0301500, and 2021ZD0300700), National Key Research and Development Program of China (Grant No. 2019YFA0705000), Natural Science Foundation of Jiangsu Province (Grant No. BK20243060), Natural Science Foundation of Guangdong Province, China (Grant No. 2025A1515011473), and National Natural Science Foundation of China (Grant Nos. 62288101, 62405382, 62293523, and 11974178).
\end{acknowledgments}

\vspace{0.5em}
\noindent\textbf{Author contributions} 

\noindent R.Y. and Y.-X.G. conceived the project. R.Y. and D.-J.G. designed the TFLN chip. Z.L. and X.C. fabricated the device. R.Y., W.Z., D.-J.G., H.K., and L.T. implemented the experiment. R.Y., W.Z., and H.K. performed the data analysis. R.Y., W.Z., and Y.-X.G. wrote the paper with input from all the authors. Y.-X.G. and S.-N.Z. supervised the project. All the authors discussed the results and contributed to the paper.

\vspace{0.5em}
\noindent\textbf{Competing interests} 

\noindent The authors declare no competing interests.
%\noindent \red{ xx are involved in developing xxx at xxx.}

\twocolumngrid
\bibliography{reference}

@article{PhysRevLett.112.120505,
  title = {Experimental Realization of Multipartite Entanglement of 60 Modes of a Quantum Optical Frequency Comb},
  author = {Chen, Moran and Menicucci, Nicolas C. and Pfister, Olivier},
  journal = {Phys. Rev. Lett.},
  volume = {112},
  issue = {12},
  pages = {120505},
  numpages = {5},
  year = {2014},
  month = {Mar},
  publisher = {American Physical Society},
  doi = {10.1103/PhysRevLett.112.120505},
  url = {https://link.aps.org/doi/10.1103/PhysRevLett.112.120505}
}

@article{roslund2014wavelength,
  title={Wavelength-multiplexed quantum networks with ultrafast frequency combs},
  author={Roslund, Jonathan and De Araujo, Renn{\'e} Medeiros and Jiang, Shifeng and Fabre, Claude and Treps, Nicolas},
  journal={Nature Photonics},
  volume={8},
  number={2},
  pages={109-112},
  year={2014},
  publisher={Nature Publishing Group UK London},
  doi = {10.1038/nphoton.2013.340},
  url = {https://doi.org/10.1038/nphoton.2013.340}
}

@article{cai2017multimode,
  title={Multimode entanglement in reconfigurable graph states using optical frequency combs},
  author={Cai, Yin and Roslund, Jonathan and Ferrini, Giulia and Arzani, Francesco and Xu, X and Fabre, Claude and Treps, Nicolas},
  journal={Nature communications},
  volume={8},
  number={1},
  pages={15645},
  year={2017},
  publisher={Nature Publishing Group UK London},
  doi = {10.1038/ncomms15645},
  url = {https://doi.org/10.1038/ncomms15645}
}

@article{jia2025continuous,
  title={Continuous-variable multipartite entanglement in an integrated microcomb},
  author={Jia, Xinyu and Zhai, Chonghao and Zhu, Xuezhi and You, Chang and Cao, Yunyun and Zhang, Xuguang and Zheng, Yun and Fu, Zhaorong and Mao, Jun and Dai, Tianxiang and others},
  volume={639},
  journal={Nature},
  pages={329-336},
  year={2025},
  publisher={Nature Publishing Group UK London},
  doi = {10.1038/s41586-025-08602-1},
  url = {https://doi.org/10.1038/s41586-025-08602-1}
}

@article{roh2025generation,
  title={Generation of three-dimensional cluster entangled state},
  author={Roh, Chan and Gwak, Geunhee and Yoon, Young-Do and Ra, Young-Sik},
  journal={Nature Photonics},
  volume={19},
  pages={526},
  year={2025},
  publisher={Nature Publishing Group UK London},
  doi = {10.1038/s41566-025-01631-2},
  url = {https://doi.org/10.1038/s41566-025-01631-2}
}

@article{
doi:10.1126/science.aad8532,
author = {Christian Reimer  and Michael Kues  and Piotr Roztocki  and Benjamin Wetzel  and Fabio Grazioso  and Brent E. Little  and Sai T. Chu  and Tudor Johnston  and Yaron Bromberg  and Lucia Caspani  and David J. Moss  and Roberto Morandotti },
title = {Generation of multiphoton entangled quantum states by means of integrated frequency combs},
journal = {Science},
volume = {351},
number = {6278},
pages = {1176-1180},
year = {2016},
doi = {10.1126/science.aad8532},
URL = {https://www.science.org/doi/abs/10.1126/science.aad8532},
}

@article{kues2017chip,
  title={On-chip generation of high-dimensional entangled quantum states and their coherent control},
  author={Kues, Michael and Reimer, Christian and Roztocki, Piotr and Cort{\'e}s, Luis Romero and Sciara, Stefania and Wetzel, Benjamin and Zhang, Yanbing and Cino, Alfonso and Chu, Sai T and Little, Brent E and others},
  journal={Nature},
  volume={546},
  number={7660},
  pages={622--626},
  year={2017},
  publisher={Nature Publishing Group UK London},
  doi = {10.1038/nature22986},
  url = {https://doi.org/10.1038/nature22986}
}

@article{reimer2019high,
  title={High-dimensional one-way quantum processing implemented on d-level cluster states},
  author={Reimer, Christian and Sciara, Stefania and Roztocki, Piotr and Islam, Mehedi and Romero Cort{\'e}s, Luis and Zhang, Yanbing and Fischer, Bennet and Loranger, S{\'e}bastien and Kashyap, Raman and Cino, Alfonso and others},
  journal={Nature Physics},
  volume={15},
  number={2},
  pages={148--153},
  year={2019},
  publisher={Nature Publishing Group UK London},
  doi = {10.1038/s41567-018-0347-x},
  url = {https://doi.org/10.1038/s41567-018-0347-x}
}

@article{mahmudlu2023fully,
  title={Fully on-chip photonic turnkey quantum source for entangled qubit/qudit state generation},
  author={Mahmudlu, Hatam and Johanning, Robert and Van Rees, Albert and Khodadad Kashi, Anahita and Epping, J{\"o}rn P and Haldar, Raktim and Boller, Klaus-J and Kues, Michael},
  journal={Nature Photonics},
  volume={17},
  number={6},
  pages={518--524},
  year={2023},
  publisher={Nature Publishing Group UK London},
  doi = {10.1038/s41566-023-01193-1},
  url = {https://doi.org/10.1038/s41566-023-01193-1}
}

@article{xie2015harnessing,
  title={Harnessing high-dimensional hyperentanglement through a biphoton frequency comb},
  author={Xie, Zhenda and Zhong, Tian and Shrestha, Sajan and Xu, XinAn and Liang, Junlin and Gong, Yan-Xiao and Bienfang, Joshua C and Restelli, Alessandro and Shapiro, Jeffrey H and Wong, Franco NC and others},
  journal={Nature Photonics},
  volume={9},
  number={8},
  pages={536--542},
  year={2015},
  publisher={Nature Publishing Group UK London},
  doi = {10.1038/nphoton.2015.110},
  url = {https://doi.org/10.1038/nphoton.2015.110}
}

@article{PhysRevA.82.013804,
  title = {Frequency-bin entangled photons},
  author = {Olislager, L. and Cussey, J. and Nguyen, A. T. and Emplit, P. and Massar, S. and Merolla, J.-M. and Huy, K. Phan},
  journal = {Phys. Rev. A},
  volume = {82},
  issue = {1},
  pages = {013804},
  numpages = {7},
  year = {2010},
  month = {Jul},
  publisher = {American Physical Society},
  doi = {10.1103/PhysRevA.82.013804},
  url = {https://link.aps.org/doi/10.1103/PhysRevA.82.013804}
}

@article{lu2023frequency,
  title={Frequency-bin photonic quantum information},
  author={Lu, Hsuan-Hao and Liscidini, Marco and Gaeta, Alexander L and Weiner, Andrew M and Lukens, Joseph M},
  journal={Optica},
  volume={10},
  number={12},
  pages={1655--1671},
  year={2023},
  publisher={Optica Publishing Group},
  doi = {10.1364/OPTICA.506096},
  url = {https://doi.org/10.1364/OPTICA.506096}
}

@article{hu2025integrated,
  title={Integrated electro-optics on thin-film lithium niobate},
  author={Hu, Yaowen and Zhu, Di and Lu, Shengyuan and Zhu, Xinrui and Song, Yunxiang and Renaud, Dylan and Assumpcao, Daniel and Cheng, Rebecca and Xin, CJ and Yeh, Matthew and others},
  journal={Nature Reviews Physics},
  volume={7},
  pages={237-254},
  year={2025},
  publisher={Nature Publishing Group UK London},
  doi = {10.1038/s42254-025-00825-5},
  url = {https://doi.org/10.1038/s42254-025-00825-5}
}

@article{dutt2024nonlinear,
  title={Nonlinear and quantum photonics using integrated optical materials},
  author={Dutt, Avik and Mohanty, Aseema and Gaeta, Alexander L and Lipson, Michal},
  journal={Nature Reviews Materials},
  volume={9},
  number={5},
  pages={321--346},
  year={2024},
  publisher={Nature Publishing Group UK London},
  doi = {10.1038/s41578-024-00668-z},
  url = {https://doi.org/10.1038/s41578-024-00668-z}
}

@article{karpinski2017bandwidth,
  title={Bandwidth manipulation of quantum light by an electro-optic time lens},
  author={Karpi{\'n}ski, Micha{\l} and Jachura, Micha{\l} and Wright, Laura J and Smith, Brian J},
  journal={Nature Photonics},
  volume={11},
  number={1},
  pages={53--57},
  year={2017},
  publisher={Nature Publishing Group UK London},
  doi = {10.1038/nphoton.2016.228},
  url = {https://doi.org/10.1038/nphoton.2016.228}
}

@article{sosnicki2023interface,
  title={Interface between picosecond and nanosecond quantum light pulses},
  author={So{\'s}nicki, Filip and Miko{\l}ajczyk, Micha{\l} and Golestani, Ali and Karpi{\'n}ski, Micha{\l}},
  journal={Nature Photonics},
  volume={17},
  number={9},
  pages={761--766},
  year={2023},
  publisher={Nature Publishing Group UK London},
  doi = {10.1038/s41566-023-01214-z},
  url = {https://doi.org/10.1038/s41566-023-01214-z}
}

@article{zhu2022spectral,
  title={Spectral control of nonclassical light pulses using an integrated thin-film lithium niobate modulator},
  author={Zhu, Di and Chen, Changchen and Yu, Mengjie and Shao, Linbo and Hu, Yaowen and Xin, CJ and Yeh, Matthew and Ghosh, Soumya and He, Lingyan and Reimer, Christian and others},
  journal={Light: Science \& Applications},
  volume={11},
  number={1},
  pages={327},
  year={2022},
  publisher={Nature Publishing Group UK London},
  doi = {10.1038/s41377-022-01029-7},
  url = {https://doi.org/10.1038/s41377-022-01029-7}
}

@article{kapoor2025electro,
  title={Electro-optic frequency shift of single photons from a quantum dot},
  author={Kapoor, Sanjay and Rodek, Aleksander and Miko{\l}ajczyk, Micha{\l} and Szuniewicz, Jerzy and So{\'s}nicki, Filip and Kazimierczuk, Tomasz and Kossacki, Piotr and Karpi{\'n}ski, Micha{\l}},
  journal={Nanophotonics},
  volume={14},
  number={11},
  pages={1775--1782},
  year={2025},
  publisher={De Gruyter},
  doi = {10.1515/nanoph-2024-0550},
  url = {https://doi.org/10.1515/nanoph-2024-0550}
}

@article{zhang2019electronically,
  title={Electronically programmable photonic molecule},
  author={Zhang, Mian and Wang, Cheng and Hu, Yaowen and Shams-Ansari, Amirhassan and Ren, Tianhao and Fan, Shanhui and Lon{\v{c}}ar, Marko},
  journal={Nature Photonics},
  volume={13},
  number={1},
  pages={36--40},
  year={2019},
  publisher={Nature Publishing Group UK London},
  doi = {10.1038/s41566-018-0317-y},
  url = {https://doi.org/10.1038/s41566-018-0317-y}
}

@article{hu2021chip,
  title={On-chip electro-optic frequency shifters and beam splitters},
  author={Hu, Yaowen and Yu, Mengjie and Zhu, Di and Sinclair, Neil and Shams-Ansari, Amirhassan and Shao, Linbo and Holzgrafe, Jeffrey and Puma, Eric and Zhang, Mian and Lon{\v{c}}ar, Marko},
  journal={Nature},
  volume={599},
  number={7886},
  pages={587--593},
  year={2021},
  publisher={Nature Publishing Group UK London},
  doi = {10.1038/s41586-021-03999-x},
  url = {https://doi.org/10.1038/s41586-021-03999-x}
}

@article{lukens2016frequency,
  title={Frequency-encoded photonic qubits for scalable quantum information processing},
  author={Lukens, Joseph M and Lougovski, Pavel},
  journal={Optica},
  volume={4},
  number={1},
  pages={8--16},
  year={2016},
  publisher={Optical Society of America},
  doi = {10.1364/OPTICA.4.000008},
  url = {https://doi.org/10.1364/OPTICA.4.000008}
}

@article{PhysRevLett.120.030502,
  title = {Electro-Optic Frequency Beam Splitters and Tritters for High-Fidelity Photonic Quantum Information Processing},
  author = {Lu, Hsuan-Hao and Lukens, Joseph M. and Peters, Nicholas A. and Odele, Ogaga D. and Leaird, Daniel E. and Weiner, Andrew M. and Lougovski, Pavel},
  journal = {Phys. Rev. Lett.},
  volume = {120},
  issue = {3},
  pages = {030502},
  numpages = {6},
  year = {2018},
  month = {Jan},
  publisher = {American Physical Society},
  doi = {10.1103/PhysRevLett.120.030502},
  url = {https://link.aps.org/doi/10.1103/PhysRevLett.120.030502}
}

@article{lu2019controlled,
  title={A controlled-NOT gate for frequency-bin qubits},
  author={Lu, Hsuan-Hao and Lukens, Joseph M and Williams, Brian P and Imany, Poolad and Peters, Nicholas A and Weiner, Andrew M and Lougovski, Pavel},
  journal={npj Quantum Information},
  volume={5},
  number={1},
  pages={24},
  year={2019},
  publisher={Nature Publishing Group UK London},
  doi = {10.1038/s41534-019-0137-z},
  url = {https://doi.org/10.1038/s41534-019-0137-z}
}

@article{PhysRevLett.94.160504,
  title = {Complementary Classical Fidelities as an Efficient Criterion for the Evaluation of Experimentally Realized Quantum Operations},
  author = {Hofmann, Holger F.},
  journal = {Phys. Rev. Lett.},
  volume = {94},
  issue = {16},
  pages = {160504},
  numpages = {4},
  year = {2005},
  month = {Apr},
  publisher = {American Physical Society},
  doi = {10.1103/PhysRevLett.94.160504},
  url = {https://link.aps.org/doi/10.1103/PhysRevLett.94.160504}
}

@article{psiquantum2025manufacturable,
  title={A manufacturable platform for photonic quantum computing},
  author = {Alexander, Koen and Benyamini, Avishai and Black, Dylan and Bonneau, Damien and Burgos, Stanley and Burridge, Ben and Cable, Hugo and Campbell, Geoff and Catalano, Gabriel and Ceballos, Alejandro and Chang, Chia-Ming and Sen Choudhury, Sourav and Chung, C. J. and Danesh, Fariba and Dauer, Tom and Davis, Michael and Dudley, Eric and Er-Xuan, Ping and Fargas, Josep and Farsi, Alessandro and Fenrich, Colleen and Frazer, Jonathan and Fukami, Masaya and Ganesan, Yogeeswaran and Gibson, Gary and Gimeno-Segovia, Mercedes and Goeldi, Sebastian and Goley, Patrick and Haislmaier, Ryan and Halimi, Sami and Hansen, Paul and Hardy, Sam and Horng, Jason and House, Matthew and Hu, Hong and Jadidi, Mehdi and Jain, Vijay and Johansson, Henrik and Jones, Thomas and Kamineni, Vimal and Kelez, Nicholas and Koustuban, Ravi and Kovall, George and Krogen, Peter and Kumar, Nikhil and Liang, Yong and LiCausi, Nicholas and Llewellyn, Dan and Lokovic, Kimberly and Lovelady, Michael and Riseti Manfrinato, Vitor and Melnichuk, Ann and Mendoza, Gabriel and Moores, Brad and Mukherjee, Shaunak and Munns, Joseph and Musalem, Francois-Xavier and Najafi, Faraz and O’Brien, Jeremy L. and Ortmann, J. Elliott and Pai, Sunil and Park, Bryan and Peng, Hsuan-Tung and Penthorne, Nicholas and Peterson, Brennan and Peterson, Gabriel and Poush, Matt and Pryde, Geoff J. and Ramprasad, Tarun and Ray, Gareth and Viejo Rodriguez, Angelita and Roxworthy, Brian and Rudolph, Terry and Saunders, Dylan J. and Shadbolt, Pete and Shah, Deesha and Bahgat Shehata, Andrea and Shin, Hyungki and Sinsky, Jeffrey and Smith, Jake and Sohn, Ben and Sohn, Young-Ik and Son, Gyeongho and Souza, Mario C. M. M. and Sparrow, Chris and Staffaroni, Matteo and Stavrakas, Camille and Sukumaran, Vijay and Tamborini, Davide and Thompson, Mark G. and Tran, Khanh and Triplett, Mark and Tung, Maryann and Veitia, Andrzej and Vert, Alexey and Vidrighin, Mihai D. and Vorobeichik, Ilya and Weigel, Peter and Wingert, Matthew and Wooding, Jamie and Zhou, Xinran},
  journal={Nature},
  volume={641},
  number={8064},
  pages={876--883},
  year={2025},
  publisher={Nature Publishing Group UK London},
  doi = {10.1038/s41586-025-08820-7},
  url = {https://doi.org/10.1038/s41586-025-08820-7}
}

@article{lomonte2021single,
  title={Single-photon detection and cryogenic reconfigurability in lithium niobate nanophotonic circuits},
  author={Lomonte, Emma and Wolff, Martin A and Beutel, Fabian and Ferrari, Simone and Schuck, Carsten and Pernice, Wolfram HP and Lenzini, Francesco},
  journal={Nature communications},
  volume={12},
  number={1},
  pages={6847},
  year={2021},
  publisher={Nature Publishing Group UK London},
  doi = {10.1038/s41467-021-27205-8},
  url = {https://doi.org/10.1038/s41467-021-27205-8}
}

@article{prencipe2023wavelength,
  title={Wavelength-sensitive superconducting single-photon detectors on thin film lithium niobate waveguides},
  author={Prencipe, Alessandro and Gyger, Samuel and Baghban, Mohammad Amin and Zichi, Julien and Zeuner, Katharina D and Lettner, Thomas and Schweickert, Lucas and Steinhauer, Stephan and Elshaari, Ali W and Gallo, Katia and others},
  journal={Nano Letters},
  volume={23},
  number={21},
  pages={9748--9752},
  year={2023},
  publisher={ACS Publications},
  doi = {10.1021/acs.nanolett.3c02324},
  url = {https://doi.org/10.1021/acs.nanolett.3c02324}
}

@article{xue_ultrabrightmultiplexed_2021,
  title = {Ultrabright Multiplexed Energy-Time-Entangled Photon Generation from Lithium Niobate on Insulator Chip},
  author = {Xue, Guang-Tai and Niu, Yun-Fei and Liu, Xiaoyue and Duan, Jia-Chen and Chen, Wenjun and Pan, Ying and Jia, Kunpeng and Wang, Xiaohan and Liu, Hua-Ying and Zhang, Yong and Xu, Ping and Zhao, Gang and Cai, Xinlun and Gong, Yan-Xiao and Hu, Xiaopeng and Xie, Zhenda and Zhu, Shining},
  year = {2021},
  journal = {Physical Review Applied},
  shortjournal = {Phys. Rev. Appl.},
  volume = {15},
  number = {6},
  pages = {064059},
  issn = {2331-7019},
  doi = {10.1103/PhysRevApplied.15.064059},
  url = {https://link.aps.org/doi/10.1103/PhysRevApplied.15.064059}
}

@article{javid_ultrabroadbandentangled_2021,
  title = {Ultrabroadband Entangled Photons on a Nanophotonic Chip},
  author = {Javid, Usman A. and Ling, Jingwei and Staffa, Jeremy and Li, Mingxiao and He, Yang and Lin, Qiang},
  year = 2021,
  month = oct,
  journal = {Physical Review Letters},
  volume = {127},
  number = {18},
  pages = {183601},
  publisher = {American Physical Society},
  doi = {10.1103/PhysRevLett.127.183601},
  url = {https://link.aps.org/doi/10.1103/PhysRevLett.127.183601}
}

@article{ma_ultrabrightquantum_2020,
  title = {Ultrabright Quantum Photon Sources on Chip},
  author = {Ma, Zhaohui and Chen, Jia-Yang and Li, Zhan and Tang, Chao and Sua, Yong Meng and Fan, Heng and Huang, Yu-Ping},
  year = 2020,
  month = dec,
  journal = {Physical Review Letters},
  volume = {125},
  number = {26},
  pages = {263602},
  publisher = {American Physical Society},
  doi = {10.1103/PhysRevLett.125.263602},
  url = {https://link.aps.org/doi/10.1103/PhysRevLett.125.263602}
}

@article{o2003demonstration,
  title={Demonstration of an all-optical quantum controlled-NOT gate},
  author={O'Brien, Jeremy L and Pryde, Geoffrey J and White, Andrew G and Ralph, Timothy C and Branning, David},
  journal={Nature},
  volume={426},
  number={6964},
  pages={264--267},
  year={2003},
  publisher={Nature Publishing Group UK London},
  doi = {10.1038/nature02054},
  url = {https://doi.org/10.1038/nature02054}
}

@article{CNOT2002,
  title = {Linear optical controlled-NOT gate in the coincidence basis},
  author = {Ralph, T. C. and Langford, N. K. and Bell, T. B. and White, A. G.},
  journal = {Phys. Rev. A},
  volume = {65},
  issue = {6},
  pages = {062324},
  numpages = {5},
  year = {2002},
  month = {Jun},
  publisher = {American Physical Society},
  doi = {10.1103/PhysRevA.65.062324},
  url = {https://link.aps.org/doi/10.1103/PhysRevA.65.062324}
}

@article{zhu2021integrated,
  title={Integrated photonics on thin-film lithium niobate},
  author={Zhu, Di and Shao, Linbo and Yu, Mengjie and Cheng, Rebecca and Desiatov, Boris and Xin, C\_J and Hu, Yaowen and Holzgrafe, Jeffrey and Ghosh, Soumya and Shams-Ansari, Amirhassan and others},
  journal={Advances in Optics and Photonics},
  volume={13},
  number={2},
  pages={242--352},
  year={2021},
  publisher={Optical Society of America},
  doi = {10.1364/AOP.411024},
  url = {https://doi.org/10.1364/AOP.411024}
}

@article{zhu2024twenty,
  title={Twenty-nine million intrinsic Q-factor monolithic microresonators on thin-film lithium niobate},
  author={Zhu, Xinrui and Hu, Yaowen and Lu, Shengyuan and Warner, Hana K and Li, Xudong and Song, Yunxiang and Magalh{\~a}es, Let{\'\i}cia and Shams-Ansari, Amirhassan and Cordaro, Andrea and Sinclair, Neil and others},
  journal={Photonics Research},
  volume={12},
  number={8},
  pages={A63--A68},
  year={2024},
  doi = {10.1364/PRJ.521172},
  url = {https://doi.org/10.1364/PRJ.521172}
}

@article{gao2022lithium,
  title={Lithium niobate microring with ultra-high Q factor above 10\^{} 8},
  author={Gao, Renhong and Yao, Ni and Guan, Jianglin and Deng, Li and Lin, Jintian and Wang, Min and Qiao, Lingling and Fang, Wei and Cheng, Ya},
  journal={Chinese Optics Letters},
  volume={20},
  number={1},
  pages={011902},
  year={2022},
  publisher={OSA},
  doi = {10.3788/COL202220.011902},
  url = {https://doi.org/10.3788/COL202220.011902}
}

@article{buddhiraju2021arbitrary,
  title={Arbitrary linear transformations for photons in the frequency synthetic dimension},
  author={Buddhiraju, Siddharth and Dutt, Avik and Minkov, Momchil and Williamson, Ian AD and Fan, Shanhui},
  journal={Nature communications},
  volume={12},
  number={1},
  pages={2401},
  year={2021},
  publisher={Nature Publishing Group UK London},
  doi = {10.1038/s41467-021-22670-7},
  url = {https://doi.org/10.1038/s41467-021-22670-7}
}

@article{herrmann2022mirror,
  title={Mirror symmetric on-chip frequency circulation of light},
  author={Herrmann, Jason F and Ansari, Vahid and Wang, Jiahui and Witmer, Jeremy D and Fan, Shanhui and Safavi-Naeini, Amir H},
  journal={Nature Photonics},
  volume={16},
  number={8},
  pages={603--608},
  year={2022},
  publisher={Nature Publishing Group UK London},
  doi = {10.1038/s41566-022-01026-7},
  url = {https://doi.org/10.1038/s41566-022-01026-7}
}

@article{hu2022high,
  title={High-efficiency and broadband on-chip electro-optic frequency comb generators},
  author={Hu, Yaowen and Yu, Mengjie and Buscaino, Brandon and Sinclair, Neil and Zhu, Di and Cheng, Rebecca and Shams-Ansari, Amirhassan and Shao, Linbo and Zhang, Mian and Kahn, Joseph M and others},
  journal={Nature photonics},
  volume={16},
  number={10},
  pages={679--685},
  year={2022},
  publisher={Nature Publishing Group UK London},
  doi = {10.1038/s41566-022-01059-y},
  url = {https://doi.org/10.1038/s41566-022-01059-y}
}

@article{jin2014chip,
  title = {On-Chip Generation and Manipulation of Entangled Photons Based on Reconfigurable Lithium-Niobate Waveguide Circuits},
  author = {Jin, H. and Liu, F. M. and Xu, P. and Xia, J. L. and Zhong, M. L. and Yuan, Y. and Zhou, J. W. and Gong, Y. X. and Wang, W. and Zhu, S. N.},
  journal = {Physical Review Letters},
  volume = {113},
  issue = {10},
  pages = {103601},
  numpages = {5},
  year = {2014},
  month = {Sep},
  publisher = {American Physical Society},
  doi = {10.1103/PhysRevLett.113.103601},
  url = {https://link.aps.org/doi/10.1103/PhysRevLett.113.103601}
}

@article{n2y3-2bmz,
  title = {On-Chip Quantum Interference between Independent Lithium Niobate-on-Insulator Photon-Pair Sources},
  author = {Chapman, Robert J. and Kuttner, Tristan and Kellner, Jost and Sabatti, Alessandra and Maeder, Andreas and Finco, Giovanni and Kaufmann, Fabian and Grange, Rachel},
  journal = {Physical Review Letters},
  volume = {134},
  issue = {22},
  pages = {223602},
  numpages = {7},
  year = {2025},
  month = {Jun},
  publisher = {American Physical Society},
  doi = {10.1103/n2y3-2bmz},
  url = {https://link.aps.org/doi/10.1103/n2y3-2bmz}
}

@article{kues2019quantum,
  title={Quantum optical microcombs},
  author={Kues, Michael and Reimer, Christian and Lukens, Joseph M and Munro, William J and Weiner, Andrew M and Moss, David J and Morandotti, Roberto},
  journal={Nature Photonics},
  volume={13},
  number={3},
  pages={170--179},
  year={2019},
  publisher={Nature Publishing Group UK London},
  doi = {10.1038/s41566-019-0363-0},
  url = {https://doi.org/10.1038/s41566-019-0363-0}
}

@article{fortier201920,
  title={20 years of developments in optical frequency comb technology and applications},
  author={Fortier, Tara and Baumann, Esther},
  journal={Communications Physics},
  volume={2},
  number={1},
  pages={153},
  year={2019},
  publisher={Nature Publishing Group UK London},
  doi = {10.1038/s42005-019-0249-y},
  url = {https://doi.org/10.1038/s42005-019-0249-y}
}

@article{parriaux2020electro,
  title={Electro-optic frequency combs},
  author={Parriaux, Alexandre and Hammani, Kamal and Millot, Guy},
  journal={Advances in Optics and Photonics},
  volume={12},
  number={1},
  pages={223--287},
  year={2020},
  publisher={Optical Society of America},
  doi = {10.1364/AOP.382052},
  url = {https://doi.org/10.1364/AOP.382052}
}

@article{CP2002,
  title = {Quantum phase gate for photonic qubits using only beam splitters and postselection},
  author = {Hofmann, Holger F. and Takeuchi, Shigeki},
  journal = {Phys. Rev. A},
  volume = {66},
  issue = {2},
  pages = {024308},
  numpages = {3},
  year = {2002},
  month = {Aug},
  publisher = {American Physical Society},
  doi = {10.1103/PhysRevA.66.024308},
  url = {https://link.aps.org/doi/10.1103/PhysRevA.66.024308}
}

@article{politi2008silica,
  title={Silica-on-silicon waveguide quantum circuits},
  author={Politi, Alberto and Cryan, Martin J and Rarity, John G and Yu, Siyuan and O'brien, Jeremy L},
  journal={Science},
  volume={320},
  number={5876},
  pages={646--649},
  year={2008},
  publisher={American Association for the Advancement of Science},
  doi = {10.1126/science.1155441},
  url = {https://www.science.org/doi/abs/10.1126/science.1155441}
}

@article{CNOTPL1,
  title = {Demonstration of a Simple Entangling Optical Gate and Its Use in Bell-State Analysis},
  author = {Langford, N. K. and Weinhold, T. J. and Prevedel, R. and Resch, K. J. and Gilchrist, A. and O'Brien, J. L. and Pryde, G. J. and White, A. G.},
  journal = {Phys. Rev. Lett.},
  volume = {95},
  issue = {21},
  pages = {210504},
  numpages = {4},
  year = {2005},
  month = {Nov},
  publisher = {American Physical Society},
  doi = {10.1103/PhysRevLett.95.210504},
  url = {https://link.aps.org/doi/10.1103/PhysRevLett.95.210504}
}

@article{CNOTPL2,
  title = {Linear Optics Controlled-Phase Gate Made Simple},
  author = {Kiesel, Nikolai and Schmid, Christian and Weber, Ulrich and Ursin, Rupert and Weinfurter, Harald},
  journal = {Phys. Rev. Lett.},
  volume = {95},
  issue = {21},
  pages = {210505},
  numpages = {4},
  year = {2005},
  month = {Nov},
  publisher = {American Physical Society},
  doi = {10.1103/PhysRevLett.95.210505},
  url = {https://link.aps.org/doi/10.1103/PhysRevLett.95.210505}
}

@article{CNOTPL3,
  title = {Demonstration of an Optical Quantum Controlled-NOT Gate without Path Interference},
  author = {Okamoto, Ryo and Hofmann, Holger F. and Takeuchi, Shigeki and Sasaki, Keiji},
  journal = {Phys. Rev. Lett.},
  volume = {95},
  issue = {21},
  pages = {210506},
  numpages = {4},
  year = {2005},
  month = {Nov},
  publisher = {American Physical Society},
  doi = {10.1103/PhysRevLett.95.210506},
  url = {https://link.aps.org/doi/10.1103/PhysRevLett.95.210506}
}

@article{CNOTmode2021,
  title = {Transverse Mode-Encoded Quantum Gate on a Silicon Photonic Chip},
  author = {Feng, Lan-Tian and Zhang, Ming and Xiong, Xiao and Liu, Di and Cheng, Yu-Jie and Jing, Fang-Ming and Qi, Xiao-Zhuo and Chen, Yang and He, De-Yong and Guo, Guo-Ping and Guo, Guang-Can and Dai, Dao-Xin and Ren, Xi-Feng},
  journal = {Phys. Rev. Lett.},
  volume = {128},
  issue = {6},
  pages = {060501},
  numpages = {6},
  year = {2022},
  month = {Feb},
  publisher = {American Physical Society},
  doi = {10.1103/PhysRevLett.128.060501},
  url = {https://link.aps.org/doi/10.1103/PhysRevLett.128.060501}
}

\clearpage
{\noindent\large\textbf{Methods}}\label{sec: method}

\vspace{0.5em}
\noindent\textbf{Device design and {fabrication}}

\noindent The DRs are designed to have 2~GHz linewidth and 12.95 GHz mode splitting. To suppress high-order mode excitation and maintain sufficient coupling rate, a 1 $\mu$m-wide waveguide and Euler curve are utilized to ensure adiabatic mode transition in the bending region. The effective radius of the Euler bend is 90~$\mu$m, and the length of the straight region is 567 $\mu$m, yielding a free spectral range of approximately 76~GHz. For high filtering efficiency and unambiguous frequency discrimination, the micro ring filters are designed to have 4 GHz linewidth and 100 GHz FSR.

The devices are fabricated using a commercially available x-cut LNOI {6-inch} wafer (NANOLN), with a {600}~nm LN thin film, a {4.7 $\mu$m} buried SiO$_2$ layer and a {500}~$\mu$m silicon substrate. First, The optical waveguides and passive devices are then patterned onto the wafer using deep ultraviolet stepper lithography. The LN thin film is then etched with a depth of {300} nm, leaving a 300~nm thick slab layer, using inductively coupled plasma reactive-ion etching process. Subsequently, a 1-$\mu$m-thick SiO$_2$ cladding layer is deposited on the top of the waveguide structures by plasma-enhanced chemical vapor deposition (PECVD). The metal electrodes are defined by patterning polymethyl methacrylate (PMMA) resist, followed by gold deposition on top of the SiO$_2$ cladding layer and lift-off.

% \vspace{0.5em}
% \noindent\textbf{Cavity-enhanced quantum light source}

% \noindent The narrow linewidth quantum light source used in our experiment is based on a monolithic doubly-resonant cavity, using a 5.3 mm-long reverse-proton-exchanged periodically poled lithium niobate (RPE-PPLN) waveguide. The cavity is formed by asymmetric coating on two facets with $94.5\%$ and $99.5\%$ reflectivity at 1560 nm, while being anti-reflective at 780 nm. The cavity is measured to have a free spectral range of 12.95 GHz and a linewidth of 202 MHz at 1560 nm. The type-0 phase-matching condition for the spontaneous parametric down-conversion process is achieved with a 15.6 $\mu$m poling period. The 780 nm pump light comes from the second harmonic generation in a 2 cm-long RPE-PPLN waveguide of an amplified 1560 nm continuous-wave laser. The temperature of RPE-PPLN cavity is stabilized to be 103.69 $^\circ$C with about 0.005 $^\circ$C fluctuation.

\end{document}